# Capturing Trojans and Irregular Satellites - the key required to unlock planetary migration

Jonathan Horner [1], F. Elliott Koch [1], and Patryk Sofia Lykawka [2]

[1] *Department of Astrophysics and Optics, School of Physics, University of New South Wales, Sydney, NSW, 2052, Australia*
[2] *Astronomy Group, Faculty of Social and Natural Sciences, Kinki University, Shinkamikosaka 228-3, Higashiosaka-shi, Osaka 577-0813, Japan*

**Summary:** It is now widely accepted that the Solar system's youth was a remarkably dynamic and chaotic time. The giant planets are understood to have migrated significant distances to reach their current locations, and evidence of that migration's chaotic influence on the Solar system abounds. The pace of that migration, and the precise distance over which it occurred, is still heavily debated. Some models of planetary migration feature systems in which the giant planets were initially in an extremely compact configuration, in which Uranus and Neptune are chaotically scattered into the outer solar system. Other models feature architectures that were initially more relaxed, and smoother, more sedate migration. To determine which of these scenarios best represents the formation of our Solar system, we must turn to the structure of the system's small body populations, in which the scars of that migration are still clearly visible.

We present the first results of a program investigating the effect of giant planet migration on the reservoirs of small bodies that existed at that time. As the planets migrate, they stir these reservoirs, scattering vast numbers of small bodies onto dynamically unstable orbits in the outer Solar system. The great majority of those bodies are rapidly removed from the system – through collisions and ejections, but a small number become captured as planetary Trojans or irregular satellites, with others being driven by the migration, leading to a significant sculpting of the asteroid belt and trans-Neptunian region.

The capture and retention efficiencies to these stable reservoirs are, to a large extent, dependent on the particular migration scenario used. Advocates of chaotic migration from an initially compact scenario argue that smoother, more sedate migration cannot explain the observed populations of Trojans and irregular satellites. Our results draw a strikingly different picture, revealing that such smooth migration is perfectly capable of reproducing the observed populations.

**Keywords:** Asteroids, trans-Neptunian objects, planetary formation, Solar system dynamics, Jupiter Trojans, Neptune Trojans, Irregular satellites

## Introduction

Evidence abounds within our Solar system that the giant planets migrated over significant distances during the final stages of their formation (e.g. [1][2][3]). Such migration is required to explain the highly excited orbits[1] of the Jovian Trojans (e.g. [4][5][6][7]), Neptunian Trojans (e.g. [8][9][10][11]), the fine structure of the asteroid belt (e.g. [12][13][14][15]), the distribution of objects in the trans-Neptunian region (e.g. [16][17][18]), and even the origin of the myriad irregular satellites of the giant planets (e.g. [19][20][21][22]).

The most obvious evidence for planetary migration in our own Solar system is the population of objects known as the Plutinos, after the first discovered, the dwarf planet Pluto[2] ([23]). These objects, trapped in 2:3 mean-motion resonance (MMR)[3] with Neptune, move on highly excited orbits ([24]), with many members, like Pluto, crossing Neptune's orbit at perihelion.

Given that the Solar system formed from a dynamically cold disk of material, featuring dust and gas on orbits with very low eccentricities and inclinations (e.g. [25][26][27]), the most likely way the Plutino population can be understood is that Neptune migrated outwards after its formation (e.g. [1][3][28]). As Neptune moved outward, it was preceded by its web of MMRs, which swept before it, trapping planetesimals and carrying them outward. Once caught in Neptune's web, the orbital inclinations and eccentricities of these objects were pumped up and they were forced ever further from the Sun, resulting in the highly excited population we observe today.

The Plutinos are not the only Solar system population that still bears the scars of planetary migration. In fact, the more objects we discover and study in our Solar system (e.g. [13][29][30][31][32]), the more evidence we find that the formation of our planetary system was a chaotic process, and that the giant planets migrated over significant distances before coming to rest at their current locations (e.g. [3][9][28]).

Despite the fact that the evidence for giant planet migration is now well accepted, the nature of that migration remains the subject of much debate. Was the migration fast or slow? Smooth or jumpy? Did the planets migrate at the same rate, or form and move at different times?[4] The answer to all these questions may lie in the evidence left behind by that migration – the distribution of the Solar system's small bodies.

In this work, we present the first preliminary results of a new dynamical study of the influence of planetary migration on the structure of the Solar system's small body reservoirs. In section two, we describe the set-up of our model, and discuss the various migration scenarios we will consider. In section three, we present preliminary results for the first of those migration scenarios. Finally, we draw our first tentative conclusions, and discuss our future work.

---

[1] Objects moving on highly excited orbits have orbits with high eccentricities and/or inclinations. Such orbits stand in stark contrast to the relatively unexcited orbits of the planets in the Solar system. In this work, we use the terms "dynamically hot" and "dynamically cold", together with talking about "excited orbits", when discussing the orbital properties of the Solar system's small body populations. A dynamically hot population is composed of objects moving on excited orbits, whilst a dynamically cold population is one in which the members move on orbits with negligible eccentricity (i.e. on near-circular orbits) and with very low inclinations (i.e. essentially within the plane of the Solar system).

[2] Following its reclassification as a dwarf planet, Pluto was added to the list of minor planets, and given the minor planet number 134340. Therefore, technically, it should be referred to as minor planet 134340 Pluto.

[3] For brevity, we hereafter use the abbreviation "MMR" for mean-motion resonance. Since the Plutinos are trapped in 2:3 MMR with Neptune, this means, on average, they complete two orbits in the time it takes the giant planet to complete three full orbits of the Sun.

[4] For examples of the many different types of migration that have been proposed to date, we direct the interested reader to e.g. [2] [7][14].

## Modelling Migration and Primordial Small Body Populations

In order to model the migration of the giant planets, and study the influence of that migration on the distribution of the Solar system's small bodies, we use the *Hybrid* integrator within the *n*-body dynamics package MERCURY ([33]). MERCURY's *Hybrid* integrator is an incredibly powerful tool for dynamical astronomy, and over the years, it has proven an invaluable tool in studies of our own Solar system (e.g. [34][35][36]), planetary habitability (e.g. [37][38][39]), planetary formation (e.g. [40][41][42]), and exoplanetary systems (e.g. [43][44][45][46]), and is easily modified in order to allow the study of planetary migration.

As described above, the various reservoirs of Solar system small bodies all bear the scars of planetary migration – and as such, it makes sense to use them as a means to study the nature of that migration. A successful migration model must explain all of the features of the Solar system's small body distributions, as well as making predictions of as yet undetected features that can be used to further test that model in future. Obviously, different types of migration (fast or slow, smooth or jumpy) would have different effects on the system's small bodies, and it should therefore be possible to determine the scale, timing, pace and smoothness of planetary migration by comparing the results of dynamical simulations to the distributions of observed Solar system small bodies.

At the time of giant planet migration, it is likely that they would have opened up gaps in the planetesimal disk, and would have cleared their immediate locale of small bodies (either by accretion or scattering). As such, whilst there would doubtless have been a population in the proto-Solar system analogous to the modern Centaurs (e.g. [47][48][49]), the great bulk of those objects would have formed elsewhere, and would simply be in the process of being scattered through the giant-planet region (much like the modern Centaurs), rather than representing a reservoir of objects that formed *in situ* that could drive the migration of the giant planets.

We therefore chose to study the impact of migration on two key test populations of small bodies. These populations would source the primordial Centaurs discussed above. The first population is analogous to the proto-Asteroid belt – a *cis*-Jovian reservoir of small bodies on initially dynamically cold orbits. The second is analogous to the proto-Edgeworth-Kuiper belt – a *trans*-Neptunian disk of material, again on dynamically cold orbits. This approach is the same as that used in our earlier work on the influence of Jupiter on the Earth's impact flux (e.g. [37][38]) and the origin of the Neptunian Trojans (e.g. [8][10]). In sum, this orbital structure, with the giant planets in pre-migration configurations and two primordial small body populations, represents the Solar system at a time when the giant planets had just formed, some ~10 Myr after the birth of the Solar system.

In our model, and following our earlier work, the planets initially migrate rapidly, asymptotically approaching their final (current) orbits (see [7][28] for more details on how this was modelled). To do this, we made use of a version of MERCURY modified to facilitate planetary migration by J. Hahn (e.g. [28]). In our simulations, Jupiter, Saturn, Uranus, and Neptune started at $a$ = 5.40, 8.60, 14.30 and 23.8 AU, respectively, values chosen to match several past studies of planetary migration [e.g. [4][5] [8][14][15][28]). We consider both fast and slow migration scenarios for the giant planets. For simplicity, we assume that Jupiter and Saturn migrate at the same rate, and that Uranus and Neptune also migrate together. Since the

Jupiter-Saturn and Uranus-Neptune pairs can evolve differently and at different times during the early evolution of the Solar system, we modelled the planet-pair evolutions by using two distinct migration timescales[5]. We therefore consider two scenarios for each planet pair. For Jupiter and Saturn, migration ceases after either 1.25 Myr (fast scenario) or 5 Myr (slow scenario), whilst for Uranus and Neptune, it ceases after 5 Myr (fast) or 20 Myr (slow). These values bracket the typical migration timescales exhibited by the four giant planets, as detailed in the past work cited above (including the most chaotic scenarios for the evolution of the outer Solar system).

For this preliminary work, we consider the influence of rapid planetary migration (i.e. Jupiter and Saturn reaching their final locations after 1.25 Myr, with Uranus and Neptune taking just 5 Myr) on a *cis*-Jovian population of test particles. That population contained test particles spread in a ring between semi-major axes of ~3.65 and ~4.85 AU from the Sun, varying in number density as a function of semi-major axis following our earlier work ([37]). From the inner edge of the disk, located three Jovian Hill radii[6] closer to the Sun than Jupiter's final orbit (i.e. beyond Jupiter's immediate zone of dynamical control once its migration has ceased, e.g. [47]), the number of objects as a function of semi-major axis rises with the square root of the semi-major axis. The outer edge of this population is located one Hill radius within the location of Jupiter's initial orbit. Rather than having a sharp cut-off, the outer edge of the disk is softened using a Cauchy function, such that the population falls to zero over a distance ten percent of the width of the main part of the disk. Mathematically, we can therefore express this distribution as follows:

$$r < r_{inner} \qquad P(r) = 0$$

$$r_{inner} < r < r_{outer} \qquad P(r) \propto r^{1/2} dr$$

$$r > r_{outer} \qquad P(r) \propto \frac{1}{1+\left(r/a\right)} \frac{dr}{a\pi}$$

where $a$ is a scaling factor set to ensure a smooth population handover across $r_{outer}$.

The orbital eccentricities of the test particles in the ring were randomly distributed between $e = 0.0$ and $e = 0.3$, while their initial inclinations were randomly distributed between $i = 0$ and 10 degrees. Our initial distributions can be seen in Fig. 1. In our future work, we will extend this disk such that the influence of migration on the full asteroid belt can be examined, but for the preliminary study, it seemed expedient to consider only the outermost (and therefore most susceptible to the influence of planetary migration) regions.

---

[5] We intend to consider the effect of different migration timescales for all four planets independently of one another in later work.

[6] The Hill radius of a planet is approximated by the relation $R_H = a_p \left(\frac{M_P}{3 M_{Sun}}\right)^{1/3}$, which is valid so long as the orbital eccentricity of the planet is low. It marks the maximum distance from the planet at which it may theoretically possess a moon, and is therefore widely used as a measure of a planet's gravitational reach. An object can be expected to be strongly perturbed if it approaches within ~ 3 $R_H$ of a given planet, and it is therefore common to consider that a given planet's gravitational reach extends to approximately that distance from its orbit. For more details, we direct the interested reader to [47].

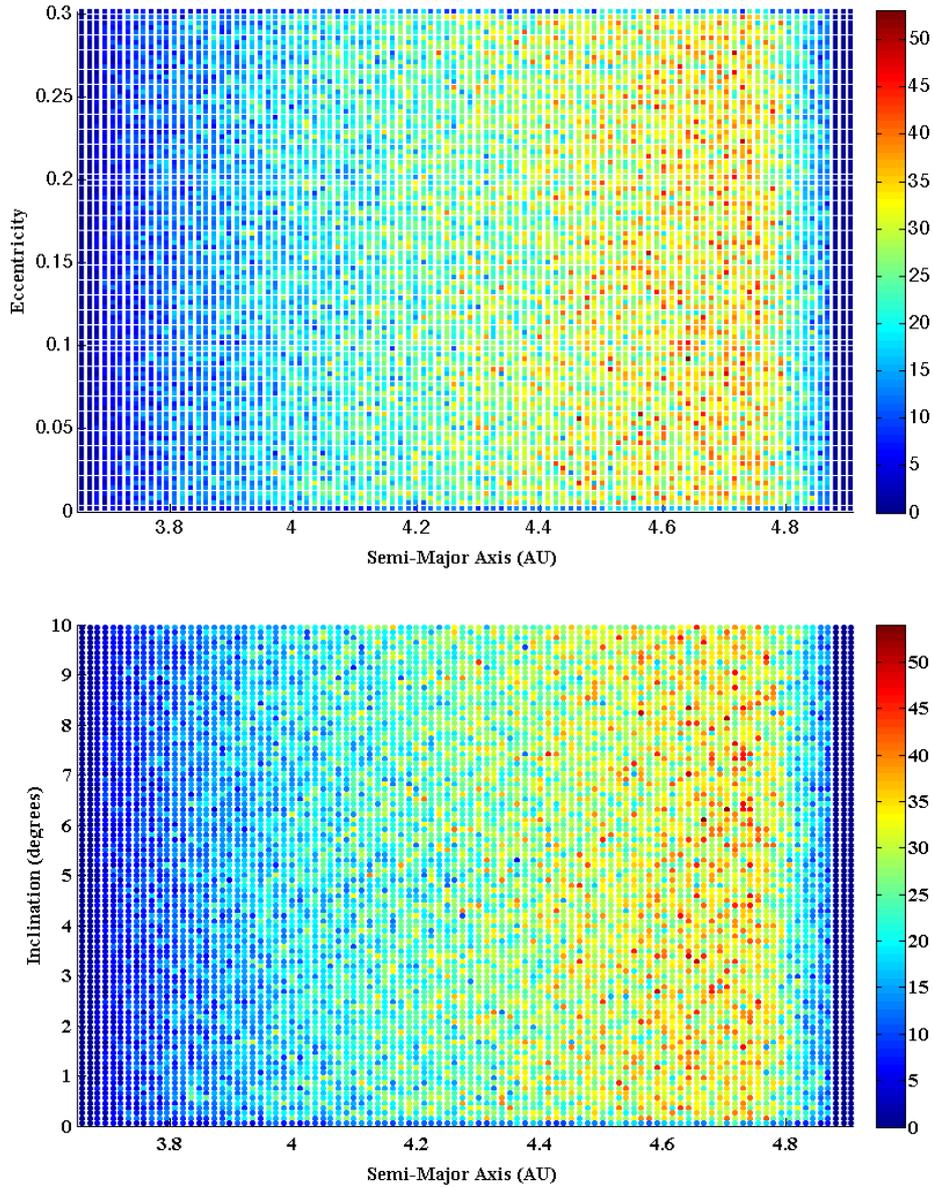

*Figure 1: The initial distribution of objects in our proto-Asteroid belt. The upper panel shows the distribution of our test particles at t = 0 in semi-major axis and eccentricity, while the lower panel shows the distribution in semi-major axis and inclination. Due to the large number of test particles considered, the data are binned, and the colour at a given a-e or a-i location shows the number of objects in that particular a-e or a-i bin.*

In total, then, we have eight distinct migration scenarios to test – two test populations, two migration types for Jupiter and Saturn, and two types for Uranus and Neptune ($2^3 = 8$). For each of these scenarios, we will eventually follow the dynamical evolution of several million test particles, initially in the chosen reservoir, until they are either ejected from the Solar system (upon reaching a heliocentric distance of 10,000 AU), or collide with one of the giant planets or the Sun. We will track the number and distribution of the particles that become captured as Trojans and irregular satellites of the various giant planets, to compare with the observed distributions in our Solar system, and will also compare the distribution of those test particles that survive until the end of our simulations to the distribution of small objects observed in the current Solar system. The simulations ran for a period of 5 Myr, coming to a halt once Uranus and Neptune had reached their final semi-major axes.

## The First Results

In total, in this preliminary work, we considered the influence of planetary migration on a population of two hundred thousand test particles distributed throughout our primordial *cis*-Jovian disk. As described above, Jupiter and Saturn migrated and reached their final orbits over a period of 1.25 Myr, while Uranus and Neptune migrated for 5 Myr before reaching their final destinations. The planets were fully mutually interacting during this time, which led to significant mutual resonant interaction – exactly what would be expected if the planets did migrate over significant distances. Fig. 2 shows the orbital elements of all test particles that survived for the full duration of our integrations, after a total of 5 Myr of dynamical evolution. For comparison, Fig. 3 shows the real distribution of over 120,000 of the numbered asteroids, as plotted in Feb. 2006. In particular, we note that, at the end of our simulations, a population of test particles are trapped in dynamically stable orbits at just under 4 AU with inclinations ranging up to ~15 degrees. This population of objects has direct analogue in the observed asteroid distribution shown in Fig. 3 – the Hilda population, trapped in 3:2 MMR with Jupiter. There are also a vast number of objects moving on dynamically unstable orbits with perihelia between the orbits of the giant planets – a transient Centaur-like population which would be expected to dynamically decay on timescales of tens of millions of years (e.g. [34]). The paucity of low-inclination objects with semi-major axes in the range 5 – 30 AU is clearly visible in the lower left-hand panel of Fig. 2. Such objects would typically be the least stable of all moving on planet-crossing orbits, since they would have a greater likelihood of experiencing close encounters with the giant planets on astronomically short timescales. We also note, in passing, that a small fraction of the test particles have evolved onto orbits with perihelia beyond the orbit of Neptune, some of which are moving on orbits that would currently place them in the Edgeworth-Kuiper belt, Scattered Disk and Detached populations (see e.g. [50]). Such objects highlight the transfer of material from one dynamically stable reservoir to another under the influence of planetary migration – a mechanism that has previously been invoked to explain the origin of the dwarf planet Ceres (which has been suggested to be an interloper from the trans-Neptunian region, rather than an object that formed in the Asteroid belt [51]).

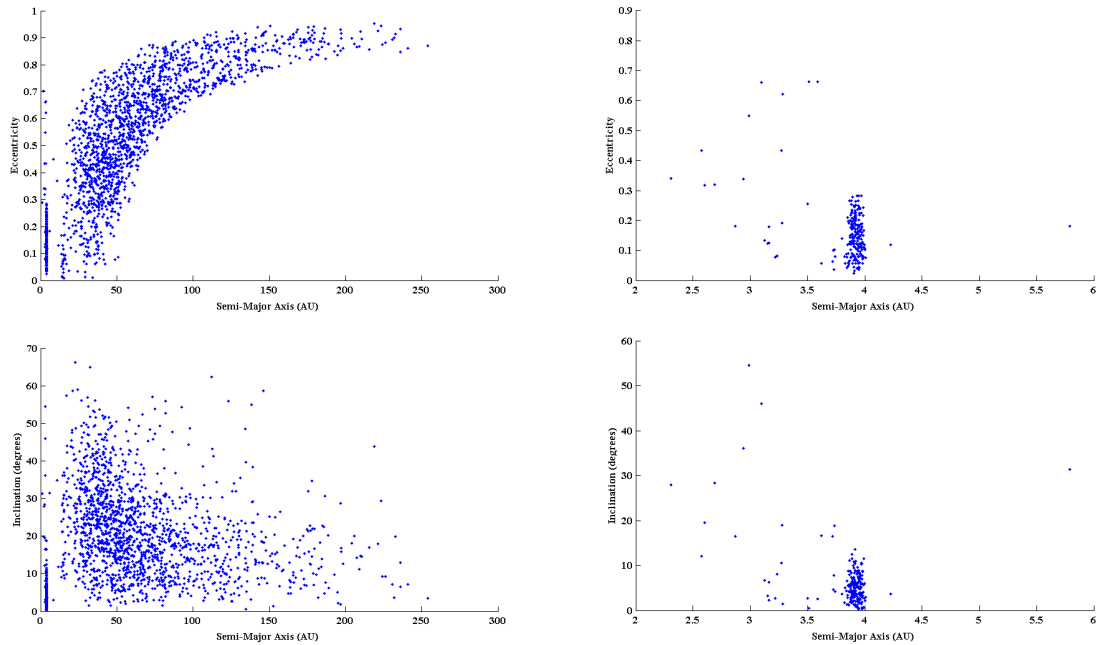

*Figure 2: The distribution of test particles that survived until the end of our integrations. The upper panels show the distribution in semi-major axis – eccentricity space, while those below show the distribution in semi-major axis – inclination space. The left hand panels show the full semi-major axis range of survivors, whilst the right show just the region between a = 2 AU and a = 6 AU. Note the strong concentration of objects at a semi-major axis of just ~3.9 AU.*

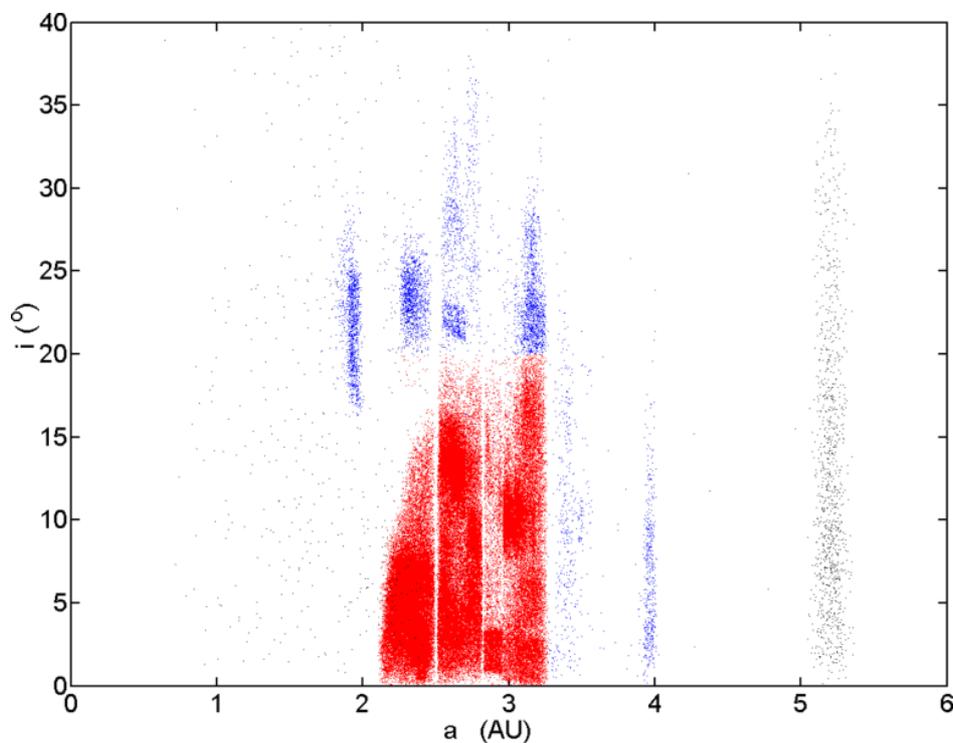

*Figure 3. The observed distribution of the Jovian Trojans and the main belt asteroids, as a function of their semi-major axis, a, and inclination, i. The figure was taken from Wikipedia (http://en.wikipedia.org/wiki/File:Main_belt_i_vs_a.png ), and was created by Piotr Deuar on 8$^{th}$ February, 2006, using orbit data for 120437 numbered minor planets, taken from the Minor Planet Center orbit database at that time. It is reproduced here under the terms of the*

Our preliminary results support those of previous studies that suggested that the Trojan capture probability for Jupiter would be of order $10^{-5}$ (e.g. [5][7]). Such a "low" capture rate is actually very promising, given that the current best estimate of the Jovian Trojan population is of order $10^6$ objects greater than 1 km in diameter, and total mass of at least ~$10^{-5}$ Earth masses ([7][52]). Given that Jupiter's migration doubtless took it through tens of Earth masses of small bodies, a capture rate of 1 in $10^5$ is certainly ample to explain the observed population.

Over the course of planetary migration, a significant fraction of the test particles that went on to be ejected from the Solar system experienced temporary captures to the various MMRs of the migrating giant planets. This is not at all unexpected – such temporary captures have been observed in dynamical simulations on many occasions (e.g. [6][7][11][53]). We present an example of such a capture event in Fig. 4.

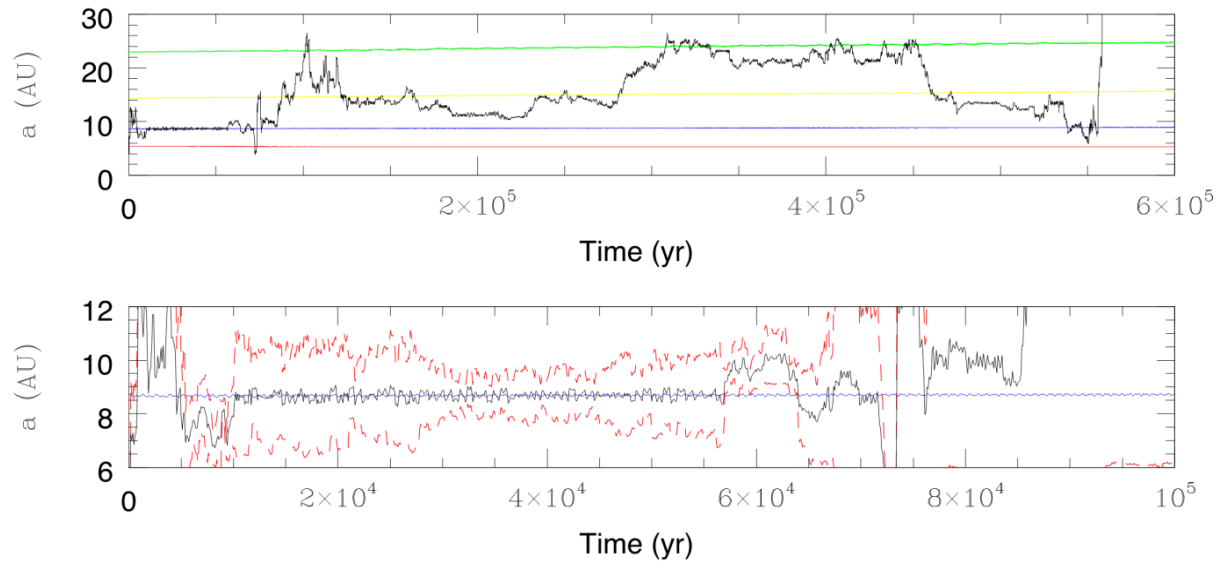

*Figure 4. The temporary capture of a test particle that originated in our proto-Asteroid belt as a Saturnian Trojan. The upper plot shows the full evolution of the semi-major axis of the object until its ejection from the Solar system, after a little less than 600 kyr. The green, yellow, blue and red lines show the evolution of the semi-major axes of Neptune, Uranus, Saturn and Jupiter, respectively, while the black line plots the evolution of the test particle's semi-major axis. The lower panel shows the first 100 kyr of the particle's evolution. Once again, the blue line shows Saturn's semi-major axis, while the black line denotes that of our test particle. The two dashed red lines show the evolution of the test particle's perihelion (lower line) and aphelion (upper line) distance.*

The evolution of the test particle followed in Fig. 4 is fairly typical of objects that escape from our proto-Asteroid belt. It rapidly evolves onto an orbit that crosses first that of Jupiter, reaching the orbit of Saturn after just a few thousand years. It is quickly captured as a temporary Saturnian Trojan (between around 10 kyr and 56 kyr), before escaping to a near-circular trans-Saturnian orbit. After another brief period as a Saturnian Trojan at the 70 kyr mark, it is ejected to a higher semi-major axis. It continues to random walk through the outer Solar system, an orbital behaviour typical of a Centaur (e.g. [48]), experiencing many short-term captures in the various MMRs of the giant planets. These captures are easily seen as periods where the semi-major axis of the object holds steady for a lengthy period – e.g. between around 340 and 350 kyr, trapped in a high order resonance just interior to the orbit of Uranus, or the lengthy period around 200 kyr, trapped in a resonant orbit between the orbits of Saturn and Uranus. Such orbital evolution, alternating between random walking and resonance sticking, is a common behaviour for dynamically unstable objects (e.g. [48][53][54]), and comes to an end when the object is ejected from the Solar system after around 560 kyr. This illustrates how objects can evolve to be captured in a different reservoir to the one in which they originated, and also highlights how challenging it is for a temporary capture to be converted into a permanent one. For this reason, although temporary captures as Trojans for short periods are common in our integrations, long-term captures are far rarer events.

## Conclusions and Future Work

The reservoirs of small bodies in our Solar system (the asteroid belt, planetary Trojan populations, the irregular satellites of the giant planets and the trans-Neptunian disk) bear the scars of the extensive migration underwent by the giant planets after their formation. Here, we present the results of a dynamical investigation of the effect of giant planet migration on objects in a proto-asteroid belt.

We model the migration of the giant planets as starting rapidly, then slowing as they approach their final destinations. We followed the dynamical evolution of twenty thousand test particles distributed through the outer reaches of a proto-asteroid belt, which stretched from an inner edge at around 3.7 AU to an outer edge just beyond 4.8 AU. We found that the great majority of the objects in our proto-asteroid belt were scattered and ejected from the Solar system. However, a significant number remained, trapped in mutual 3:2 mean-motion resonance with Jupiter. The distribution of these test particles in semi-major axis – inclination space was essentially identical to that of the Hilda population observed today.

The test particles that escaped from the proto-asteroid belt typically moved onto Jupiter-crossing orbits, and then evolved as typical Centaurs for tens or hundreds of thousands of years until a close encounter with one of the giant planets flung them out of the Solar system forever. A significant fraction of the test particles collided with the giant planets – a result that is not unsurprising, since it is reasonable to expect that the giant planets would still be accreting large amounts of material as they migrate.

The majority of the test particles experienced repeated short-term captures in the many MMRs of the giant planets. Such captures were generally short, however, with only a tiny fraction of those objects experiencing long-term capture. Our results support earlier work, finding that the long-term capture probability as a Jovian Trojan is of order $10^{-5}$. Although such a capture probability sounds punishingly low, it is actually more than ample to explain the currently observed Jovian Trojan population, since Jupiter's migration would have caused it to encounter tens of Earth masses of small bodies before it reached its final destination.

In the future, we will continue our studies by performing seven further simulations, each examining either a different migration scenario for the giant planets, or a different source reservoir for the small bodies involved (a proto-Edgeworth-Kuiper belt). By the end of our work, we should be able to better constrain the true nature and pace of planetary migration, using the known populations of small bodies in our Solar system today as a guide to disentangle the events that occurred over four billion years ago.

## Acknowledgements

The authors wish to thank J. Hahn, who made available the version of MERCURY, modified to enable planetary migration, used in this work. The authors also wish to thank the two referees of this paper, Dr. Paul Francis and Dr. Rob Wittenmyer, for their comments, which helped to improve the flow and clarity of our article.